# A REVIEW OF PROGRESS FOR COMPONENT BASED SOFTWARE COST ESTIMATION FROM 1965 TO 2023

**Adjunct Faculty and PhD Scholar (CS) Muhammad Nadeem**

Department of Computer Science, FCIT, University of Punjab Lahore, Pakistan

Department of Computer Science, Institute of Computing, BZU Multan, Pakistan

**Prof. Dr. Humaira Afzal**

Department of Computer Science, Institute of Computing, BZU Multan, Pakistan

**Associate Prof. Dr. Muhammad. Idrees**

Department of Computer Science and Engineering, University of Engineering and Technology Lahore, Narowal Campus, Pakistan

**Assistant Prof. Dr. Sajid Iqbal**

Department of Computer Science, Institute of Computing, BZU Multan, Pakistan

**Prof. Dr. M. Rafiq Asim**

Department of Computer Science, COMSATS University Islamabad, Lahore Campus. Pakistan.

## ABSTRACT

Component Based Software Engineering (CBSE) is used to develop software from Commercial Off the Shelf Components (COTs) with minimum cost and time. Component Based Software Cost Estimation (CBSCE) is an important pre-development activity for the successful planning and cost estimation of Components-Based Software Development (CBSD) that saves cost and time. Many researchers are putting their efforts to propose and then develop a CBSCE model. This motivates to review research work and history of CBSCE from 1965 to 2023. The scope of this research also, to some extent, includes auxiliary the review of all the research work done in the areas such as CBSE, CBSCE, Component Based Software Metrics, COTs, component based process models to cover all the areas of CBSD under CBSE either to answer or to provide pointers for the answers to the questions of this area easily. Internet based search methodology has been used to review the available and published literature. This paper may also classify available literature of this area into its sub areas such as component selection, quality with chronological contribution of the researchers and pictorial presentation of its history. Thus this research paper may serve as a common source of information for the concerned researchers.

**INTRODUCTION**

As per the famous quotation of Aeschylus, "The one knowing what is profitable, and not the man knowing many things, is wise". In this scenario cost estimation before starting software development for the sake of computerized business is a profitable activity. So much so that wrong estimates may lead to even complete failure of not only software but also business involved. SCE is not an easy or simple process because of many reasons defined by (Heemstra, 1992). There are two types of estimation in this context. One is software estimation tools and other one is project management tools. The project management tools started 10 years ahead of software cost estimation. The project management tools started around 1960 and the SCE started before 1970. But significant SCE started in 1965 when System Development Corporation (SDC) studied a lot of attributes of many projects (Nelson, 1966). It may also be mentioned here that there are two types of software cost estimations with respect to commercial software estimation tools in logic (Capers, 2013). These are macro and micro estimations. Macro estimation is top to bottom estimation and micro estimation is bottom to top estimations. The difference between the two is that macro estimation deals with complete project estimation and micro estimation deals with activity wise estimation. As per the investigation done by (Barry, 2000) all the available types of SCE techniques have been categorized into six major categories such as model-based, experienced-based, learning oriented, dynamic based, regression-based and composite. SCE marketplace was formed by the investigators of the top notch leading organizations such as IBM, Hughes, RCA, TRW, and the U.S. Air Force. The research of these organizations laid the foundations for the development of commercial cost estimation tools. As far as the motivation of the researchers is concerned, the authors of (Sarah, 2007) have given findings from the organized review of the literature for the motivation of software engineers in software engineering. According to (Boehm, 1981), motivation in software engineering has the single largest impact on the productivity of the practitioners and quality management (Mcconnell, 1998) as well.

An Excellent exploration study has been conducted by (Maras, 2012) that are about 15 years history of CBSE Symposium for its impact on research community.

The next section II shows the review process. In section III, analysis of existing study has done; results of the research provided in section IV; Validity of research given in section V; Summary of this paper is given in section VI; Limitations and future work is given in section VII; references are given n section VIII.

# REVIEW PROCESS

## INCLUSION CRITERIA

Following are the major inclusion areas of the 58 years (from 1965 to 2023) history to search and then review research articles where software components are involved:

1. CBSCE
2. CBSE
3. CBSD

**Table 1** Research questions with main purposes

| S.# | Research Question | Main Purpose |
|---|---|---|
| 1 | What, Why and When Componentization is/was started and needed? | To support CBSCE researchers for better know-how of the Componentization. |
| 2 | To what extent are CBSCE researchers aware of the potential of need of researchers on CBSD? | To identify the most important CBSCE publications from the beginning of Componentization to date. |
| 3 | Which are the years of the least development of CBSD and CBSCE? | To find the start of CBSCE area and the contribution of the investigators. |
| 4 | Which are the years of the maximum development of CBSD and CBSCE? | To evaluate the maximum progress of CBSCE area and the contribution of the researchers. |
| 5 | What is the yearly growth rate of the CBSE, CBSCE and CBSD? | To evaluate and review the progress of CBSCE area and the contribution of the researchers in chronological order. |

| 6 | How many categories of the research papers published to date of CBSE, CBSCE and CBSD? | To identify the trends, strengths and weaknesses of the categories of CBSE, CBSCE and CBSD |
|---|---|---|
| 7 | How many researchers who have a long term interest in CBSCE? | To evaluate the susceptibility of the SCE research of CBSCE. |
| 8 | What are the major explored CBSCE research issues and Why? | To identify the trends, strengths and weaknesses of the categories of CBSE, CBSCE and CBSD for reducing software development time and cost. |
| 9 | How many review studies were conducted for SCE from 1965 to 2023? | To identify the statistical information about the potential of CBSE, CBSD and financial trends. |
| 10 | How many number of papers published for the history of CBSE and CBSD? | To identify the trends, strengths and weaknesses of the categories of CBSE, CBSCE and CBSD |
| 11 | How many research papers have been published on CBSCE area? | To identify the research work that has been already done and to find limitations of further research. |
| 12 | How many research articles have been published regarding in-order history of CBSE, CBSCE and CBSD? | To identify and assesses whats have already been done, from where to start and who is who and what is what was and now regarding CBSE, CBSCE and CBSD. |

| 13 | Why and how much fruitful it is to summarize the chronological history of CBSCE, CBSE and CBD in a single paper? | So that novice researchers and practitioners of CBSE, CBSCE and CBSD can easily know the past and current findings and future research topics of the area. |

## SOURCES OF RESEARCH PAPERS

The main source of relevant papers is the internet. We also search the required papers from different types of world level reputed journals such as relevant journals of IEEE, ACM, Springer, ISI Thomson, Google Scholar citations, NASA cost estimation, Higher Education Commission (HEC) of Pakistan Recognized Journals, CBSE and CBSD conferences and symposiums. This research work mainly hinges on Mclory, Bohem Barry, Albright, Magne Jorgensen, Martin Shepperd, Capers Jones besides many others etc.

## CLASSIFICATION OF RESEARCH PAPERS

Research papers are classified into following areas as per the process model of CBSE, CBSCE and CBSD:

**Table 2** Subareas of CBSE

| S.#. | Subareas |
| --- | --- |
| 1 | A great idea buy, dont build that later converted to CBSD |
| 2 | A review for accuracy of effort and schedule estimation. |
| 3 | Components Testing |
| 4 | CBSD as new age of software development. |
| 5 | Components Risks |

| 6 | Validation process of CBSD |
|---|---|
| 7 | Components Importance |
| 8 | Challenges of CBSE and CBSD |
| 9 | Motivations for software engineers. |
| 10 | Issues and challenges in SCE of CBSD |
| 11 | Effort estimation in CBSD |
| 12 | Finding Required Components |
| 13 | CBSCE |
| 14 | Components Selection |
| 15 | A basic model of CBSCE of only the maintenance cost |
| 16 | Component Based Development (CBD) |
| 17 | Effort models that are belong to CBSD models |
| 18 | Framework for a SCE Model for Software Modification |
| 19 | Components Customization |
| 20 | Components Reliability Estimation |
| 21 | Compatibility ratio of the new components |
| 22 | Components Complexity. |
| 23 | Component based software static and dynamic metrics |
| 24 | Component based software process model |
| 25 | Component-Based Framework for Software Reusability |
| 26 | Components packaging |
| 27 | COTs or COCOTS |
| 28 | Components Standardizations |
| 29 | Optimization of Components |
| 30 | Components Quality |

# SYSTEMATIC REVIEW

One of the valuable study on the history of SCE is performed by (A. A. Syed, 2012) in detail but the focus of research of this paper is an attempt to summarize the history of CBSCE using CBSE. Later on, 40 years history of SCE methods and techniques from 1965 to 2005 is provided in [34]. There are different SCE methods such as Algorithmic Cost Methods (Models) and Expert Judgment and Machine Learning Methods (Attarzadeh, 2010). Review research methodology is used to validate this research work.

The idea for CBSCE started in 1969 by Mcilroy (Bauer, 1968) and has become renowned after giving the terminologies as componentization and product line software engineering problems.

The idea of software reuse was initiated in 1968 by Mcllroys in the NATO Software Engineering Conference. This conference also renowned as the birthplace of the software engineering filed (Bauer, 1968).

Development of large industrial software systems with highest reliability and availability requirements results in a great cost. Thats why different companies started to develop such costly system by reusing already developed components. But the work of (Elaine, 1998) proved that those components that a company planned to reuse must be tested first to avoid a huge loss. Many studies (Jenkins, 1984, Phan, 1990, Heemstra, 1992, Lederer, 1995, Bergeron, 1992, Sauer, 2003) have been done for the accuracy of effort and schedule, projects that were completed was over estimated. The authors of the research (Morisio, 2002) analysed 15 projects at the Flight Dynamics Division at the Goddard Space Flight Center of NASA. They concluded that COTs based development process is quite different from traditional software development. There are many risks and challenges identified in (Padmal, 2003). After that new challenges and problems faced by CBSE and CBSD were explained and examined by (Ivica, 2003, Sahra, 2003). Another excellent job for the enhancement of SCE research via a systematic review of the published research work has been done by (Magne, 2007). They reviewed 304 journal research papers of 76 different journals. Some parameters that have their impact for the effort estimation in CBSD are in (Puneet, 2009). After this, the extended version of the UML (Unified Modelling Language) means RE-UML (Requirements Engineering - UML) given by (Mahmood, 2009) enable the system analyst to find the correct required components that satisfied the requirements of the stakeholders. Another technique of components selection and optimization for SCE was proposed by (Sedigh, 2005) using graph based model for CBSD. The investigation done by (Raghu, 1997) reveals the importance to Know-How, Know-Why and Know-What to understand the history of any area affairs to CBSD. The author explored two components of knowledge such as Know-Why and Know-What. Know-How knowledge is learning by doing by (Arrow, 1962; Dutton, 1985; Argote, 1990). Know-Why knowledge is the learning by studying by [29]. Know-What knowledge component is learning by using by (Rosenberg, 1982; Von Hippel, 1984; Karone, 1993).

The authors of (Klk, et al., 2008) have given the assessment criteria to authenticate process models. The research work of (Agarwal et al., 2001) has given a comparative analysis of various methods and techniques to find software cost. The authors of (Dijkstra, 1968) emphasized the importance of the organization of the program in 1968.

According to them organization of program is as important as the development of the program.. The author of (James, 1984) discussed the Draco approach to software development from reusable components. Then in 1986 a great idea "buy, dont build" idea was given by (Brooks, 1987) that was later converted to CBSD. After this the investigators of (CHRIS et al., 1987) discussed their to mention the importance of software reuse in (CHARLES, 1992) that results in the reduction of software cost and development time and ensure good quality. Again in 1992, the authors highlighted the importance of packaging components before storing these components into components repository, increase quality and then thus productivity in (Basili et al., 1992). As per (Moløkken et al., 2004) and (Clements, 1995) being gave details of historical, logical and technical shifting from subroutines to subsystems,, subsystems to components are available. During talk the author of (Victor, 1995) discussed a lot with managing director Angela Burgess. But following are the important findings of his research:

1. Companies have been running on Institution but Institution is not always right.

2. Technology transfer is built into the experience factor.

The research work of (Albert and Jayesh, 1995) found the causes of inaccurate SCE in 1995. The researchers of (Elaine, 1998) has given that it is necessary to test a component before reuse to avoid a big loss. Resultantly the research work of (Mikio, 1998) provided the pictorial representation for new era of software development

. The world renowned cost estimation scientist and the author of (Boehm, 1999) research conducted for managing software productivity and reuse in 1999 for reusability. Then in February 2000, a nice research work on software risk management has been done by (Ropponen, 2000). They identified different software components risks as scheduling and timing risks etc. Then software development cost estimation approaches was provided by (Boehm, 2000) in detail. The authors of (Douglas and Laurence, 2000) again discussed and reviewed the core reasons due to which developing countries such as Zambia does not get the advantages of CBSD. An overview of different types of SCE techniques and models were given by (Javier, 2000). After that in October 2000, a validation process of CBSD for the estimation of the size of the software has been performed (Javier, 2000). In May, 2001, the importance of the COTS was determined by the finding of (Boehm, 1999) and (Victor and Boehm, 2001) that more than 99% of computer instructions get from COTS products. The research work done by (Shachter, 1986; Moløkken, 2003) the authors provided the SCE summary by the review of the review of software effort estimation. In 2004, the author of (Larsson, 2004) research highlighted the possibility of developing component technologies that provide mechanism for predicting quality attributes of software system. His main finding is the classification of different quality attributes that can found directly from the properties of the components and for that need more detail like usage profile or architecture. The authors investigated 112 projects of Chinese software project benchmarking dataset in (Yang et al., 2008). Different types of issues and challenges in SCE are found and discussed in November 2008 in the research work done by (Zaid, 2008). The research work done in (Khatibi et al., 2010) contributed a lot to find out the reasons of software failure due to wrong and doubtful cost estimation pessimistically and optimistically in 2010.They recommend the project managers to study and apply suitable cost estimation model as per the nature and history of the projects. The research work of (Kaur and Mann, 2010) provides the overview of the CBSE in May 2010.  As per the systematic review of CBSCE, the research work done by (Nadeem et al., 2010) in December 2010 is the first paper on CBSCE. It is the first and initial step in CBSCE taken by (Nadeem et al., 2010). According to this research work, it is found a number of parameters involved in CBSCE, gave levelling, importance and contribution of these parameters. One of the main parameters is standardization of components either in standalone or distributive environment had highest significance communality percentage of variance. The research work of (Xiaotie and Miao, 2011) about the summarization of SCE discussed the importance of SCE as a key filed for effective cost management. The authors also elaborated three most important cost estimation methods and gave detailed overview of COCOMO2 module. The research conducted by the authors of (Jovan and Dragan, 2012) in January 2012, found that cost and size estimation of a software system is a main challenge. A basic model of CBSCE of only the maintenance cost of the software project on the basis of COCOMO in June 2012 is given by (Siddhi and Rajpoot, 2012). Their basic cost estimation model consists of three parameters such as, Annual Change Traffic, Non-Technical factors and Technical Development cost of Component Based Software which affect the maintenance cost of Component Based Software. Now another wonderful step was taken by (Khan et al., 2012) that is the development of state of the art Component Based Development (CBD) models namely V Model, Y Model, W Model, X Model and ELCM in 2012 An Integrated Component-Based Development (ICBD) Life cycle and model was given by (Rekaby and Osama, 2012) in November 2012. They also provided different levels of reusability in CBSD. As per the outcomes and the results his new model, it reduced the effort of the projects up to 40% within three months of its use. It is a big achievement in component based software cost estimation. Following quotations given by Boehm Barry in (Boehm, 2015)

are also very important for the recognition of SCE in October 9, 2002:

i.     Poor management can increase software costs more rapidly than any other factor.

ii.    Fix specification errors early. To fix later, they will cost:

* 520% more at design stage

* 1,000% more at coding

* 2,000% more at unit test

* 20,000% more at delivery

Another important review for Software effort estimation approaches and risk analysis was conducted by (Poonam, 2012). They described the different effort models belonging to CBSD.

Framework for Developing a SCE Model for Software Modification had been developed by (Hathaichanok, et al., 2012).

In August 2012, selection and customization framework for CBSD was proposed by (Lata et al., 2012). As per their research output, they said that main problems of software development on the basis of CBSD are components selection and components customization to satisfy the requirements of the software to be developed. Reliability Estimation of Component-Based Software methodology was proposed by (Singh and Tomar, 2012) in 2012. Another

Algorithm for Component Selection with X Model was given by (Pradeep and Gill, 2013) for CBSD. The authors of (Kumar et al., 2010) came up with the idea of the factors to find the optimal components as per the requirements of the clients. In 1986 investigators of (IEEE, 1990; Boehm, 1986)] discussed the factors named as performance, size, reliability, fault tolerance, time and complexity in 1990 and a spiral model of software development and enhancement was also given and elaborated in (Boehm, 1986).

Now in 2013, latest model for reliability estimation of CBSD was given by (Aditya and Pradeep, 2013). This model also estimates the impact factor of individual factors. ,

A new automation tool to find the compatibility ratio of the new components for the new software project was developed in May 2013 (Nishant and Dhawaleswar, 2013). Basics of CBSE, its process and different types of metrics such as cost and complexity etc. are discussed nicely by (Divya et al., 2013). The summary of the metrics, models and tools for SCE is given by (Soumyabrata et al., 2013) in September 2013. These tools included ACEIT (Automated Cost Estimating Integrated Tools), Agile COCOMO II, CSE (Center for Software Engineering Tools), COOLSoft, COSMOS, Cosatr, PMPal, r2ESTIMATOR, Taasc Estimator and SEER. The researchers of (Sagayaraj and Poovizhi, 2013) highlighted the component based software metrics.

A generic model of SCE using a hybrid approach was developed by (Lalit et al., 2014) in February 2014. Component based Software Process Model named as Elite Plus was given (Lata and Neena, 2014). Another remarkable achievement was performed by (Jahanzaib and Aasia, 2014) giving effort estimation method on the basis of lifecycle in CBSE. A Component-Based Framework for Software reusability was also provided by (Adnan et al. 2014). Reusability helps a lot for an organization to reduce the software development cost using CBSD.

The authors of (Zahid Khan and Khan, 2014) given a central value based software repository. It can help the customers in a better way in searching best possible reusable component in 2014. The contribution of these researchers helps a lot to increase the maximum level of components reusability. Significant factors for reliability estimation of component based software systems was given by (Kirti et al., 2014).

Again in the year of 2014, a comprehensive study of component based complexity metrics was done by (Pooja and Rajender, 2014). The authors (Gnanasankaran and Iyakutti, 2014) highlighted the definition, features and characteristics of COTs in details in 2014. They discussed the six-step methodology proposed by the Minkiewicz (Minkiewicz, 2005). It shows the necessary processes to occur for the development of the COTs. Now in 2014, one of the main research works by the authors of (Magne, 2014) played a major role for CBSD cost estimates. They strongly emphasized to understand and well communicate the meaning of an effort or cost estimation to prevent planning and budgeting mistakes. The authors of (Magne, 2014) also recommend conclusively that "the meanings of effort estimates are understood and communicated using a probability–based terminology". One of the significant factors of (Nadeem et al., 2010) is the standardization of components in standalone or distributed environment. This research was conducted in 2010. After four years in 2014, the research work of (Lahon and Sharma, 2014) also reproved the work done by (Nadeem et al., 2010).

According to the research work of (Tanwar et al., 2014) in 2014, traditional software development is not capable of meeting the requirements because it puts forward the already built in advantages of software quality, development productivity and finally total cost of the software to be developed.

Now very recently in the year 2015, analysis of Capability Maturity Model Integrated (CMMI) level 5 for developers was performed by the (Corinne et al., 2015). As they found that there is no any metrics exist for the performance of the software, so they used proxy for top quality software. The authors of (Arti et al., 2015) given the software component quality model for CBSE on the basis of IEEE 1061 and ISO 9126. They also found idea of software quality and its relationship with cost under the strategies of quality management.

Now another most important A Review on Software Sizing for Project Estimation was conducted by the authors of (Eric, et al., 2015). They have given the birds eye view of the traditional and second generation software sizing methods. The researcher presented the evaluations of the quality of the software components using tools such as Edrawl Tool and Understand 2.6 Tool in (Gagan, 2015).

Next review analysis was conducted for software quality assurance in CBSD by (Abeer et al., 2015) in the same year 2015 for the confirmation of the quality of components. According to them, a core reason behind this review analysis was that qualitative component is not providing a guarantee without proper evaluation and confirmation regarding quality of components.

Another wonderful research was done by (Latika, 2015) for the testing cost estimation of the quality of software. According to them not only reliability but also reusability, correctness, maintainability as well as assure the quality and productivity of the software need a metrics using an effective software testing cost measurement technique within budget.

An excellent hand book has been written by the valuable authors of NASA in (Conde et al., 2015). The authors of this cost estimation handbook of NASA provided the summary of the process for developing cost estimates. According to them, cost estimation process is a key to make a process successful within budget. The NASA needs cost estimates for the formulation used for CBSD projects and/or other projects. Much of the details for every phase are given in this book but the authors of this paper are giving only the cost estimation techniques and one parametric cost estimation technique is selected for our methodology for research.

This cost estimating process is used to find the estimates on the basis of statistical relationships among historical cost, system performance characteristics etc. One of the most significant benefit of this parametric cost estimating technique is that it gives quick estimates easily.

The authors of (Deepa and Saurabh, 2015) have analysed different factors for SCE in the then situation in 2015. The main objective of the authors was to find what types of SCE methods were used by software industry those days? They reviewed a questionnaire in different large, medium and small software companies. After the analysis, they divided their findings into following five factors such as SCE technique, nature of the project, training individuals in cost estimation, review process for the estimated software cost and risk buffer.

According to the results mentioned by the SPSS analysis using Symmetric measure and Chi square test on the parameter w. r. t. company size, the impacts of the factors aforementioned is directly proportional to the size of the company. It further notified that for example if we consider risk buffer factor then its value was greater for large company size as compared to intermediate and/or small size companies. Now in September 2015, an excellent research work was done by the authors (Latika, 2015). They proposed another quality assurance model for CBSD to eliminate quality defects in first phase and to make the software a high quality component based system.

They reviewed different component based technologies and also the characteristics of these technologies. Their proposed quality assurance model covered component quality and its interactions. They have a plan to test their proposed model in component based software industry as well.

The research work done in (Denning, 2015) highlighted the security concern for the software system. A core reason for this research was the addition of 7,937 vulnerabilities to The National Vulnerability Database (NVD) of USA, up from 5,174 in 2013. This is approximately 22 per day, or almost one every hour. Out of these vulnerabilities, 1,912 (24%) were marked as high severity and 7,243 (91%) high or medium.7 simply but, they cannot be ignored. So they found that the cost of the vulnerabilities is varied from few hundred to hundred thousand dollars. Thus keeping in view, the cost for the secure software must be kept in plan to tackle the vulnerabilities.

The authors of (Mittas and Mamalikidis, 2015) presented a wonderful framework of the errors of SCE methods for visualization and statistical comparisons using a StatRec toolkit. Their suggested framework provides good help to provide planning for decision making for software development.

A verity of algorithm approaches for efficient retrieval of components repositories are discussed by (Bawa1 and Iqbaldeep, 2016). One of the major findings of (Bawa1 and Iqbaldeep, 2016) is that robustness of components can be increased by the classification of components. The authors of (Sharanjit, 2016) discussed about the components based software development cycle. They have also highlighted the importance of the selection of quality components to reduce cost as well as software development time. Different types of software cost estimation techniques have been reviewed by (Shivangi and Umesh, 2016). They also gave benefits and drawbacks of various software cost estimation techniques. There are two main types of software cost estimation techniques such as algorithmic and non-algorithmic techniques. A major output of their research is that the software engineers should use the combination of different cost estimation methods for efficient estates. It may also be mentioned here that there is even now no single method to find the best software cost for all projects. Software cost estimation using dynamic reusability estimation model as per reusability at design level is proposed by (Mangayarkarasi and Selvarani, 2017) using mathematical analysis. This model also gives the feedback to know the required effort and cost to develop a new system. A comparative analysis of different techniques for software cost estimation was given in detail along with pros and cons of each technique by (Zaffar et al., 2017). These researches also concluded that its not easy to choose accurate method due to different scenarios of the projects to be developed. They also recommended many key things such as dont stuck on one software cost estimation method, for cost and time frame use different methods/models. Another research has been conducted by (Javed and Faisal, 2017) to analyse and optimize cost estimation model COCOMO-II. This analysis and optimization was done for enterprise level software in Pakistan. This research work highlighted one of the major finding that COCOMO-II is the best software cost estimation model for enterprise level in Pakistan.

A very useful, present, past and future review research work with respect to future systems and software challenges and especially maintainability was presented very nicely by the great author of software cost estimation in (Boehm, 2017). The author highlighted the increasing cost of software maintenance and technical debts very scientifically and technically.

The researchers of (Tribhuvan et al., 2018; Antonio et al., 2019; Carlos et al., 2020; Shachi et al., 2021; Maedeh et al., 2022; Shaima et al., 2023 ) also proposed SCE and CBSCE algorithms and effort estimation.

The sequential summary of these techniques is given below in table 3:

## 58 YEARS CHRONOLOGICAL ORDER OF CBSCE

**Table 3** 58 Years chronological order of main

Contributions/achievements of CBSCE

| S. # | Year | Main Contribution(s) |
|---|---|---|
| 1 | 1962 | Learning by doing given |
| 2 | 1966 | Statistical history of 169 software projects |
| 3 | 1968 | Emphasized the importance of an organization for the development of program. |
| 4 | 1969 | An idea for CBSCE started in 1969 by Mcilroy and he became renown after giving the terminologies as componentization and product line software. |
| 5 | 1981 | Found motivation in software engineering |
| 6 | 1982 | Learning by using |
| 7 | 1984 | Novel product concepts from lead users and Segmenting users by experience |

| 8 | 1984 | A study was conducted for accuracy of effort and schedule estimation |
| 9 | 1984 | Discussed Draco approach for the construction of software systems from reusable software parts. |
| 10 | 1985 | Know-Why learning by studying was given. |
| 11 | 1986 | An influence diagram as subsets of software metrics given [22]. |
| 12 | 1986 | A spiral model for software development and enhancement was given and discussed |
| 13 | 1987 | A great idea buy, dont build that later converted to CBSD was given. |
| 14 | 1987 | Albrechts function points effort estimation model was validated |
| 15 | 1990 | An Integrated Resource Planning Perspective Model was given. |
| 16 | 1990 | Know-How learning was given. |
| 17 | 1990 | Different types of CBSCE factors were found and discussed. |
| 18 | 1992 | SCE difficulties are addressed and discussed |
| 19 | 1992 | A review was conducted for effort accuracy and schedule estimates. |
| 20 | 1992 | A pilot study was done on estimation of Information Systems Development |

|     |      | Efforts. |
| --- | ---- | -------- |
| 21  | 1992 | Importance of software reuse discussed |
| 22  | 1992 | Importance of packaging components mentioned |
| 23  | 1993 | Approaches to innovation in modern wind energy technology was elaborated by Know-What. |
| 24  | 1995 | Inaccurate software development cost estimates reasons found |
| 25  | 1995 | Historical, logical and technical shifting from subroutines to subsystems, subsystem to components details was elaborated. |
| 26  | 1995 | Foundation of the software engineering as an engineering process rather than manufacturing process provided. |

**Continue … Table 3**

| S. # | Year | Main Contribution(s) |
| ---- | ---- | -------------------- |
| 27   | 1995 | Found inaccurate SCE reasons |
| 28   | 1997 | Introduced efficient components of knowledge |
| 29   | 1998 | Emphasized developed testing components before reuse |
| 30   | 1998 |      |
| 31   | 1998 | Practical discovery of tested components usage. |

| 32 | 1998 | Changed from traditional to CBSD software development |
| 33 | 1999 | Managing software productivity and reusability with Eight different types of critical success factors for reuse found |
| 34 | 2000 | A general idea of variety of SCE techniques and models |
| 35 | 2000 | Six software components risks were identified |
| 36 | 2000 | An overview of different types of SCE techniques and models were given via a review. |
| 37 | 2000 | Core reasons that developing countries did not get the advantages of CBSD |
| 38 | 2000 | Validation process of CBSD for the estimation of the size of the software has been performed |
| 39 | 2001 | The importance of the COTS |
| 40 | 2001 | Introduction searched based software engineering a new area |
| 41 | 2001 | A comparative analysis of various methods and techniques to find software cost was performed. |
| 42 | 2002 | COTS-based software development, processes and open issues was given |
| 43 | 2003 | The status of information technology project management was given to |
| | | determine the accuracy of effort estimates. |
| 44 | 2003 | Risks and challenges of CBSD was identified and discussed. |
| 45 | 2003 | Challenges and problems faced by CBSE & CBSD was researched. |
| 46 | 2003 | Found Present challenges in cost and quality management of CBS |
| 47 | 2003 | SCE summary was provided by the review of the review of software effort estimation |
| 48 | 2004 | Review for SCE was performed |
| 49 | 2004 | Mechanism for predicting quality attributes of software system was provided |
| 52 | 2005 | Another technique of components selection and optimization for SCE was proposed |
| 51 | 2005 | Six steps to a successful COTS Implementation were given. |
| 52 | 2007 | Motivations for software engineers were discussed |
| 53 | 2007 | An organized analysis of software development cost estimation studies was presented |
| 54 | 2008 | Assessment criteria and Pros and cons of many SCE techniques from 1965 to 2005 were discussed |
| 55 | 2008 | An inspection of 112 projects for Chinese |
| | | software benchmarking dataset performed. |
| 56 | 2008 | Issues and challenges in SCE were found and discussed |
| 57 | 2009 | Identification of effort estimation parameters in CBSD |
| 58 | 2009 | Extended version of Unified Modelling Language given |
| 59 | 2010 | A Novel Algorithmic Cost Estimation Model Based on Soft Computing Technique given. |
| 60 | 2010 | Reasons of software failure due to wrong estimation pessimistically and optimistically found. |
| 61 | 2010 | CBSE overview with the components selection and reuse |
| 62 | 2010 | The number of parameters involved in CBSCE and then given the levelling, significance and contribution of these parameters were given by the systematic review of the literature and the review of project managers of the software industry |
| 63 | 2010 | New most favourable process for component selection was given to select best components. |
| 64 | 2010 | Systematic literature reviews in software engineering performed. |
| 65 | 2011 | Summary of SCE was |

| S. # | Year | Main Contribution(s) |
|---|---|---|
| | | discussed |
| 66 | 2012 | Fifteen years history of CBSE Symposium was elaborated. |
| 67 | 2012 | A renowned historic cost estimation techniques review was conducted and discussed |
| 68 | 2012 | Estimation of a software cost was found and discussed |

**Continue … Table 3**

| S. # | Year | Main Contribution(s) |
|---|---|---|
| 69 | 2012 | A basic model of CBSCE of only maintenance cost of the software on the basis of COCOMO was given. |
| 70 | 2012 | A state of the art CBD models developed. |
| 71 | 2012 | An Integrated Component-Based Development Life cycle and model was given. |
| 72 | 2012 | Six CBSD effort models given |
| 73 | 2012 | A structure as a framework for developing a SCE model for software modification developed |
| 74 | 2012 | Selection and customization framework for CBSD was proposed. |
| 75 | 2012 | Reliability Estimation of Component-Based Software methodology was proposed. |
| 76 | 2013 | Analysis of brief description of the history of software estimation |

| | | tools presented |
|---|---|---|
| 77 | 2013 | Another latest method for component selection was given. |
| 78 | 2013 | A new model for reliability estimation of component-based software was given. |
| 79 | 2013 | A new automation tool to find the compatibility ratio of the new components for the new software project was developed. |
| 80 | 2013 | Basics of CBSE, its process and types of metrics discussed |
| 81 | 2013 | A summary of metrics, models and tools for SCE was given |
| 82 | 2013 | Component based software metrics was discussed |
| 83 | 2014 | A generic model of SCE using a hybrid approach developed. |
| 84 | 2014 | An elite plus CBD model was presented and discussed. |
| 85 | 2014 | A very new CBSE lifecycle model called Circular Process Model was given. |
| 86 | 2014 | A Component-Based Framework for Software Reusability was provided to reduce software cost |
| 87 | 2014 | A Value Based Software Repository was developed to enhance software reusability. |
| 88 | 2014 | Many important features for reliability estimation |
| | | for CBSD were analyzed and provided. |
| 89 | 2014 | A comprehensive study of component based complexity metrics was done. |
| 90 | 2014 | The definition features and characteristics of COTs were elaborated in software. |
| 91 | 2014 | The strong emphasize to understand and communicate meaning of an effort/cost estimation to prevent planning/budgeting mistakes taken |
| 92 | 2014 | After four years in 2014, this research work has also proved that CBSE facing many challenges because of the lack of the standardizations that was the finding of [61]. |
| 93 | 2014 | An evaluation on optimized COCOTS model in CBSE approach was conducted. |
| 94 | 2015 | Poor management effects on cost |
| 95 | 2015 | Found that traditional software development is not capable to meet the requirements was researched |
| 96 | 2015 | The software component quality model for CBSE on the basis of IEEE 1061 and ISO 9126 was given. |
| 97 | 2015 | A Review on software sizing for project estimation was conducted |

| S. # | Year | Main Contribution(s) |
|---|---|---|
| 98 | 2015 | Evaluations of the quality of the software components presented. |
| 99 | 2015 | A review analysis was conducted for software quality assurance in CBSD by [97] |
| 100 | 2015 | A cost estimation metrics was proposed |
| 101 | 2015 | A very excellent hand book has been written by the valuable authors of NASA |
| 102 | 2015 | Analysis of Impacts of different factors on SCE now a days |
| 103 | 2015 | Another CBSD quality assurance model proposed. |
| 104 | 2015 | The security concern for the software system highlighted |
| 105 | 2015 | A wonderful framework of the errors of SCE methods for visualization and statistical comparisons was presented |
| 106 | 2015 | Research-paper recommender systems were proposed |
| 107 | 2016 | Efficient retrieval of components repositories algorithm discussed. |

**Continue … Table 3**

| S. # | Year | Main Contribution(s) |
|---|---|---|
| 108 | 2016 | Highlighted the importance of selection of quality components to reduce cost |

| | | as well as software development time. |
|---|---|---|
| 109 | 2016 | Different types of software cost estimation techniques reviewed. |
| 110 | 2017 | Software cost estimation using dynamic reusability estimation model has been proposed using mathematical analysis. |
| 111 | 2017 | A comparative analysis of different techniques for SCE was given with pros and cons |
| 112 | 2017 | Highlighted major finding that COCOMO-II is the best software cost estimation model for enterprise level in Pakistan. |
| 113 | 2017 | The increasing cost of software maintenance and technical debts have been elaborated very scientifically and technically |
| 114 | 2018 | SCE using Environment Adaptation Method |
| 115 | 2019 | Estimating costs of multi-component enterprise applications |
| | 2020 | Software Development Effort Estimation |
| 116 | 2021 | Learning Component Size Distributions for SCE |
| | | A Novel framework to improve analogy-based SCE |
| 117 | 2022 | Software Project Effort Estimation |

**Table 4** Summary of Publications from 1962 to July 2023 w. r .t. Yearly Research Output

| S. # | Published Year | No. of CBSD Publications | No. of Years |
|---|---|---|---|
| 1 | 1962, 1966, 1968, 1969, 1981, 1982, 1985, 1993, 1997, 1999, 2002, 2011 | 1 | 12 |
| 2 | 1986, 1987, 2004, 2005, 2007, 2009 | 2 | 6 |
| 3 | 1984, 1990, 2001, 2008 | 3 | 4 |
| 4 | 1995, 1998 | 4 | 2 |
| 5 | 1992, 2000, 2003 | 5 | 3 |
| 6 | 2010 | 6 | 1 |
| 7 | 2013 | 7 | 1 |
| 8 | 2012 | 10 | 1 |
| 9 | 2014 | 11 | 1 |
| 10 | 2015 | 13 | 1 |
| 11 | 2016 | 15 | 1 |
| 12 | 2017 | 20 | 1 |

| | 13 | 2018-2023 | 6 | 6 |

Table 4 shows that the researchers published 1 research paper of CBSCE and CBSD area in 12 different year that are given in above table, 2 research papers of CBSCE and CBSD area in 6 different year, 3 research papers of CBSCE and CBSD area in 4 different year and so on.

But according to this table, 2017 is that year in which most of the research output produced by the investigators globally that is 20 numbers of publications in one year on CBSD and CBSE basis. It is also observed that speed of research work increased from 2010 and significantly increasing year by year. This progress shows that researchers may become successful to develop a reliable and practicable CBSCE model in coming years soon.

Following Fig.1 shows the graphical presentation of the Table-4 in term of No. of CBSD Publications with respect to years:

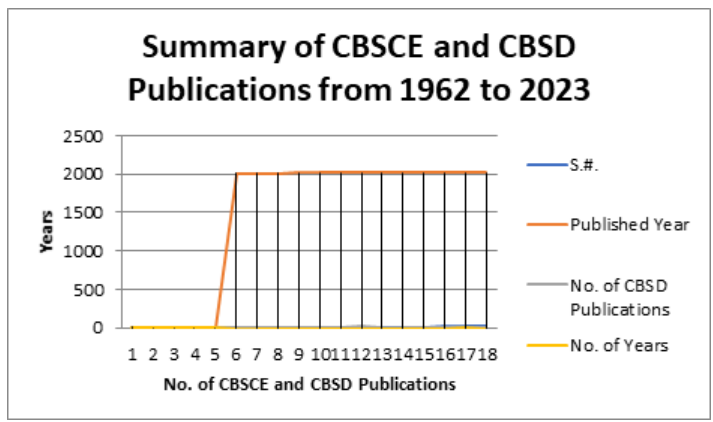

**Fig. 1**: Summary of CBSCE and CBSD Publications from 1962 to 2017 Separately

this rate of publications considerably grew from 2010 to 2015. Three years namely 2012, 2014 and 2015 were excellent number of years as per the research work output of this area. The publications of these years being crossed the single digits of 10, 11 and 12 respectively. This key and keen analysis can easily be seen in figures 1 above in this paper.

**RESULTS OF THE RESEARCH**

**What, Why and When Componentization is/was started and needed? (RQ1)**

We found that componentization has been started in 1969 by Mcilroy. It was and now needed to develop cheaper software projects by using already developed software components in minimum time. CBSD is playing its role in fastest software projects development with reliability, functionality and greater productivity etc.

**To what extent are CBSCE researchers aware of the potential of need of researchers on CBSD? (RQ2)**

The researchers of the area were well aware about components and finding its cost since 1967 [1]. The extent of the potential for the need of the researcher can easily be estimated that in 1999 it was found by the [48] that 99% of the code instructions come from already developed components.

**Which are the years of the least development of CBSD and CBSCE? (RQ3)**

As per the findings of Table3, in 12 years named as1962, 1966, 1968, 1969, 1981, 1982, 1985, 1993, 1997, 1999, 2002, 2011, one research paper has published in the CBSCE. But there were many papers that published on CBSD, SCE etc.

**Which are the years of the maximum development of CBSD and CBSCE? (RQ4)**

Again as per the findings of Table 3, maximum development of CBSD and CBSCE has been done in the year 2015 when 13 papers were published. But major developments were started in 2010 when 6 papers were published in one year.

**What is the yearly growth rate of the CBSE, CBSCE and CBSD? (RQ5)**

Growth rate of CBSE, CBSCE and CBSD year wise is given in Table 3.

**How many categories of the research papers published to date of CBSE, CBSCE and CBSD? (RQ6)**

We used review and/systematic review of the published research work in related area of CBSCE in CBSE. Following Table-4 shows that we have found 35 different subareas of CBSD, CBSCE and CBSE. This table also shows the references of the research articles and total number of research papers of the relevant subareas.

**ANALYSIS**

The systematic review, 58 years chronological order summary, classification of sub areas of CBSE, CBSCE, CBSD, Total number of published research papers per year and collective and the number of published papers as per the classification of subareas provided a broader view of the properties, strengths and weaknesses of CBSE, CBSCE and CBSD. This broad and narrow view becomes the source of starting point to investigate the findings in details as per the authors perspective along with suggestions and findings for further proposals and development of CBSCE techniques and/or models. Furthermore, table 2 provides not only the history of step by step growth of CBSCE but it also shows major findings and/or contributions with complete references. Again table 3 highlights the yearly progress trends of the publications of CBSCE and CBSD. It gives a clear cut scholarly eye sight that there was a very slow rate of research work publications from 1962 to 1968 but

**Table 5** CBSE, CBSD and CBSCE subareas and its total number of papers (1965-2017)

| S. # | Subareas | Papers |
|---|---|---|
| 1 | Componentization | 1 |
| 2 | Software construction, Reusability and productivity | 3 |
| 3 | A great idea buy, dont build that later converted to CBSD | 1 |
| 4 | Accuracy and effort estimation | 13 |
| 5 | Packaging Components | 2 |
| 6 | Composing software with CBSD | 1 |
| 7 | Components Testing | 2 |
| 8 | CBSD as new age | 1 |
| 9 | Components Risks | 2 |
| 10 | CBSD Validation process | 1 |
| 11 | Components Importance | 1 |
| 12 | CBSE & CBSD Challenge | 2 |
| 13 | software engineers Motivations | 1 |
| 14 | Issues and challenges in SCE | 1 |
| 15 | Effort and cost estimation techniques in CBSD | 3 |
| 16 | Finding Required Components | 1 |

| 17 | CBSCE | 1 |
|---|---|---|
| 18 | Components Selection | 2 |
| 19 | 15 years history of CBSE | 1 |
| 20 | Cost and size estimation | 2 |
| 21 | A basic model of CBSCE [64 | 2 |
| 22 | CBD | 1 |
| 23 | CBSD Effort models | 1 |
| 24 | Framework for a SCE Model for Software Modification | 1 |
| 25 | Components Customization | 1 |
| 26 | Components Reliability Estimation | 4 |
| 27 | Compatibility ratio of the new components | 1 |
| 28 | Components Complexity | 3 |
| 29 | Component based metrics | 1 |
| 30 | Component based software process model | 3 |
| 31 | Component-Based Framework for Software Reusability | 2 |
| 32 | COTs | 4 |
| 33 | Components Standardizations | 1 |
| 34 | Components Optimization | 2 |
| 35 | Components Quality | 5 |
| Total | | 74 |

**CBSD, CBSE and CBSCE Subareas =$\sum_{1}^{n} RP$**

Where, RP= Research Papers and n = 74.

**How many researchers who have a long term interest in CBSCE? (RQ7)**

There are many renown globally for long term interest in SCE in general and CBSCE such as Mclory, Dr. Bohem Barry, Magne Jorgensen etc.

**What are the major explored CBSCE research issues and Why? (RQ8)**

These are Components Reliability Estimation, Components Testing, Components Risks, Challenges of CBSE and CBSD, Components Selection etc. Because these are playing vital role for the CBSD that leads to CBSCE as well.

**How many review studies were conducted for SCE from 1965 to 2017? (RQ9)**

We found 13 different review and review studies as given in table3.

**How many number of papers published for the history of CBSE and CBSD? (RQ10)**

We found three such papers in total. One of 15 years history of CBSE symposium [8] and next are 40 years history of SCE (Boehm and Valerdi, 2008), a short history of SCE tools (Capers, 2013).

**How many research papers have been published on CBSCE area? (RQ11)**

As far as our review is concerned, we found the (Nelson, 1966) in 1966, (Xiaotie et al., 2011) that were published in 2010 and has been cited in 4 different journals on CBSCE particular area and (Rekaby and Osama, 2012), (Jahanzaib and Aasia, 2014) as well. Many other scientists have also published their incomplete research work.

**How many research articles have been published regarding in-order history of CBSE, CBSCE and CBSD? (RQ12)**

These are 119 but other 3 papers used just for the validity of review methodology.

**Why and how much fruitful it is to summarize the chronological history of CBSCE, CBSE and CBD in a single paper? (RQ13)** CBSE and CBSD are using now days to develop software projects very fast in less time and cost with greater productivity, reliability and efficiency as well. So many researchers and practitioners have been jumping into this area of research to its advantages. The research is beneficial for the software developers and as well as clients. In addition to this many latest software developments tools such as Java and Dot Net are based on CBSD. So there is a crying need of the relevant professionals to learn, apply and then devise and revise the improved model etc for the better cost estimation.

**How?**

A chronological history of CBSCE, CBSE and CBD in a single paper is helpful for the recent or novice researcher to get almost all the relevant stuff of CBSD, CBSE and CSBCE in a single platform.

**VALIDITY OF RESEARCH**

Review methodology is used to validate this research by review of the research papers published in National and/or International authenticated research journals or conferences.

The referenced research work that is searched via Internet is the backbone of this paper. The validation of publications thus found is beyond any doubt because of authenticity of their sources. In addition to this, the research work of (Mark and Baryan, 2001) that is the research on a new area of software engineering named as searched based software engineering is another bold logical and research oriented validation of this research methodology. In addition to this review identified 119 published research papers of the area during 58 years in-order of significant contributions/achievements of CBSCE given in table 2 above has also validated this research. In spite of this, figures 1 are also provided above on the basis of the sequential order of CBSCE. Another important validation of this review is the progress of the publications of CBSCE given in table 3. This table shows that number of publications of CBSCE is increasing year by year rapidly but gradually. This too validates research work of this paper.

## SUMMARY

This manuscript analysed the CBSCE, CBSD and CBSE research papers published in different National and International reputed journals and tried to search other papers such as metrics for components selection, component complexity, components integrity, quality of components and maintenance cost of components etc. through comprehensive online review. This historical review identified 119 CBSCE, CBSD and component based SCE metrics paper in various reputed global level journals. We have also studied 13 different types review studies for accuracy of effort and schedule estimation. We suggest the following on the basis of the chronological history and the research contributions of different authors in the areas of CBSCE, CBSD and CBSE:

1. We didnt find any research paper that provided us the relevant research papers, reports and books of this area at a single platform that we did now.
2. Increase the breadth and depth of research for relevant studies.
3. Need to conduct more research studies to develop a CBSCE model.
4. Increase the awareness of the researchers, relevant academicians and professionals in software industry about the benefits of CBSD and CBSE. After that the importance of CBSCE should be highlighted to reduce failure rate of software projects and to quickly develop cheaper software.
5. This paper also provides a wonderful sequence of overview and outcomes of the research work. So this paper may act as great sources, suitable container for beginners/practitioners to know-what, know-how and know-why of CBSD and CBSE that may lead towards proper CBSCE model.

## LIMITATIONS AND FUTURE WORK

## LIMITATIONS

A detailed review of the literature of CBSCE is conducted by us to summarize the history of CBSCE. One of the main points is that CBSE and CBSCE is a new area and many researchers and practitioners are globally working and adding to this area day-by-day. Their objective is to develop cheap software project in minimum time. So we cannot assure to have summarized the complete history of the CBSCE. Next limitation is that we have been successful to work only on the papers that we have published ourselves e.g. (Nadeem et al., 2010) and the papers or literature available online given in reference section. So we could not work on the published work that is available in libraries/research labs due to lack of resources.

## FUTURE WORK

There is an urgent need to search the unaddressed literature globally by requesting and/or visiting to CBSE research journals/libraries/research and development organizations and then update the relevant history. Secondly it is also required to find the comparative analysis of different types of CBSCE techniques for the novice researchers so that they can think to improve or propose new technique for CBSCE.

## ACKNOWLEDGMENT:

The authors would like to thank all the research bodies such as IET, IEEE, Springer, Research gate. They are also grateful for a number of International research journals such as IET Software, IEEE Software etc. for published qualitative literature regularly that provide a strong foundation for this paper.